# Photon-gated spin transistor

Fan Li, Cheng Song*, Bin Cui, Jingjing Peng, Youdi Gu, Guangyue Wang, and Feng Pan

Key Laboratory of Advanced Materials (MOE), School of Materials Science and Engineering, Tsinghua University, Beijing 100084, China

E-mail: songcheng@mail.tsinghua.edu.cn

Abstract

Spin-polarized field-effect transistor (spin-FET), where a dielectric layer is generally employed for the electrical gating as the traditional FET, stands out as a seminal spintronic device under the miniaturization trend of electronics. It would be fundamentally transformative if optical gating was used for spin-FET. We report a new type of spin-polarized field-effect transistor (spin-FET) with optical gating, which is fabricated by partial exposure of the $(La,Sr)MnO_3$ channel to light-emitting diode (LED) light. The manipulation of the channel conductivity is ascribed to the enhanced scattering of the spin-polarized current by photon-excited antiparallel aligned spins. And the photon-gated spin-FET shows strong light power dependence and reproducible enhancement of resistance under light illumination, indicting well-defined conductivity cycling features. Our finding would enrich the concept of spin-FET and promote the use of optical means in spintronics for low power consumption and ultrafast data processing.

Since the first field-effect transistor (FET) was fabricated by Shockley in 1952,[1] FET has undergone a tremendous progress and promoted extensive applications in electronics, with the emergence of novel devices of different types.[2–7] Spin-polarized field-effect transistor (spin-FET) proposed by Datta and Das in 1990,[8] as a seminal spintronic device, has become an intriguing field of research with the potential for the solution to the influence of the quantum effect under the trend towards miniaturization in electronics. It has drawn extensive attention for the realization of spin-FET features in several material systems and device configurations.[9–12] However, as the main method for the manipulation of the performance in spin-FETs, the effective electrical control of the normal conductive channel is seriously limited by the screening effect, while the realization in devices with two dimensional electron gas is restricted by spin injection, which presents a dilemma in front of researchers.

Optical manipulation of magnetism, considered as an ultrafast manner, has demonstrated great potential in device applications for ultrafast data processing and storage.[13,14] To achieve optical manipulation of magnetic ordering, much work has been done, providing the possibility to manipulate spins in an optical way, such as the optical excitation of magnetic precession,[15] the all-optical magnetization reversal,[16] and the photon-induced phase transition and spin transition,[17–19] confirming the optical control of the spins. So far, gate operation of spin-FET features has been realized by electric field[20–22], magnetic field[11] and interface resonance[23,24], which can be considered as an electrical-field effect in nature. It would be fundamentally transformative if the optical method was used as a field-like manner for gate operation in spin devices, proposing a concept of the photon-gated spin transistor, which would further promote advances in ultrafast spintronics. In addition to the superiority of fast switching, the photon-gated spin transistor shows advantages of no leakage problem and low power consumption, when compared to the commonly used electrical counterpart. In the following, we experimentally design a photon-gated spin transistor and demonstrate an optical control of spin transport, with half-metallic La$_{1/2}$Sr$_{1/2}$MnO$_3$ (LSMO) serving as the channel.

High-quality LSMO epitaxial films with the thickness of 20 unit cell were prepared on SrTiO$_3$ substrates. Films were then patterned into Hall bars with a width of 100 μm and transport measurements were performed at 10 K in a physical property measurement system with an in-situ LED (light-emitting diode) set over the sample as a light source. Before investigating the features of the photon-manipulated spin transistor, we demonstrate the capability of spin manipulation via the optical method. The channel resistance, i.e. the magnetoresistance (MR), as a function of the in-plane magnetic field along the channel was measured when LEDs of different colors (red, yellow, green, and blue) were adopted with the power fixed at 3 mW. **Figure 1**a shows MR curves measured with light of different colors or without light. No matter what color of the light is, the two curves gained by sweeping the field from positive to negative and from negative to positive separate from each other and the window enclosed by these two curves becomes larger under light illumination, in comparison to the case without light. In view of the nature of anisotropic magnetoresistance, the enlargement of the MR window reflects the enhanced saturated field,[25] which can be attributed to the photon-induced spin transition as reported in our previous work.[19] That is to say, the photon-excited down spins replace the major up spins in LSMO and thus the trend for an antiparallel alignment of up spins and down spins makes the saturation of magnetization more difficult. In consequence, it is concluded that an effective spin transition occurs under all the light of different colors. In order to see the influence of the light color more clearly, Figure 1b summarizes the relative area of the MR window, which is calculated by the integral of the area enclosed by the two curves of different scanning direction, with the area under no illumination set to 100%. All the area of the MR window measured under light illumination is more than 160%, indicating an enlargement of the window. It is obvious that as the photon energy increases, that is, as the color varies from red to blue, the relative area increases to the largest value under the illumination of the blue light, meaning that the most effective optical manipulation of spins can be achieved under the blue light among these four kind of light.

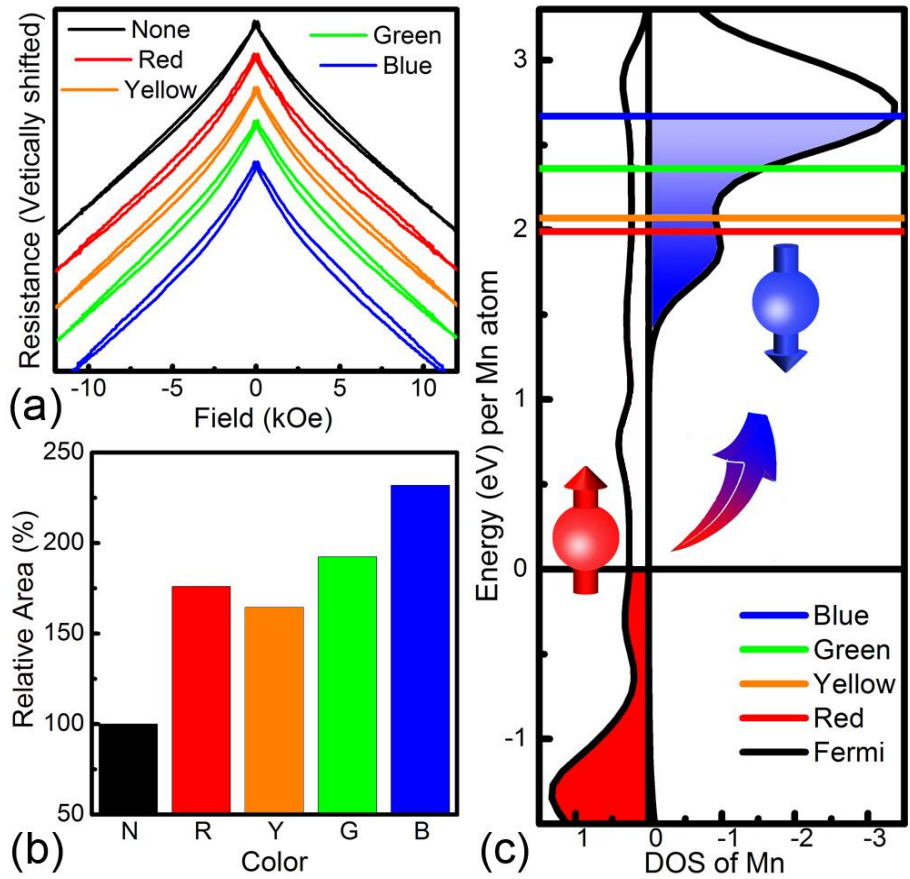

**Figure 1.** a) Dependence of MR loops on light color. Curves are vertically shifted for clarify. b) The relative integral area of the window in MR loops under different light with the value under no light as a reference. c) DOS of Mn atom in $La_{1/2}Sr_{1/2}MnO_3$ with the photon-induced spin transition from up spins to down spins shown by the arrows. Colored horizontal lines denote the energy level excited by the light of different colors.

To investigate the mechanism lying under the color dependent efficiency of the optical manipulation, first-principles calculations were carried out on the basis of a 2×2×4 supercell with the stoichiometry of $La_{1/2}Sr_{1/2}MnO_3$ with density of states (DOS) of a Mn atom depicted in Figure 1c. As a typical half-metallic electronic structure, there is a large spin polarization near the Fermi level, i.e. the Fermi level is fully occupied by up spins, while the DOS of down spins is almost zero. Note that as the energy increases, the DOS of the down spins over the Fermi level increases rapidly and would exceed that of the up spins. In this view, when the sample is under light illumination of different colors, electrons at the Fermi level would jump to different energy levels corresponding to the light energy, as marked by the horizontal colored lines in Figure 1c. Since the DOS of the down spins (the minor spins) is distinctly larger than that of the up spins (the major spins) at the excitation level for all the visible light, most of the up spins at the Fermi level turn to down spins during the light excitation, while only a small number maintain the initial up spins. As a consequence, under the illumination of visible light appear many down spins, which tend to show an antiparallel alignment to the major up spins and make the saturation more difficult, leading to the photon-induced enlargement of the MR window in Figure 1a. It is noteworthy that the largest (smallest) value of the spin-down DOS is gained at the level of the blue (yellow) light, while the spin-up DOS varies little as the light color changes. This implies that the degree of the photon-induced spin transition from major spins to minor spins is largest (smallest) under the illumination of the blue (yellow) light. Consequently, the relative integral area of the MR window reaches the maximum and the minimum for the blue light and the yellow light, respectively. In this view, it is the asymmetry of the spin DOS above the Fermi level and the effective excitation under visible light that allow LSMO to show an effective photon-induced spin transition, which is unavailable in either non-ferromagnetic materials or transition metals. So far, we have proved the possible optical manipulation of spins in LSMO in both experiments and calculations, offering the basis for the optical gate operation in the photon-gated spin transistor, and the blue light was selected as the light source in the following experiments for the largest degree of the photon-induced spin transition.

Based on the discussion above, we designed a new type of spin transistor with an optical gate, as illustrated by the optical image of the device in the inset of **Figure 2**a. Opaque insulating tapes (the blue part) were used to cover part of the channel (1000×100 μm$^2$), leaving a 300 μm long area under light exposure. In this way, the light illumination would cause a difference in spin polarization between the exposed area and the other area. Thus a gate operation of the proposed photon-gated spin transistor is expected, considering the photon-induced resistance increase due to the enhanced scattering of the spin-polarized current when it goes through the photon-gated area.

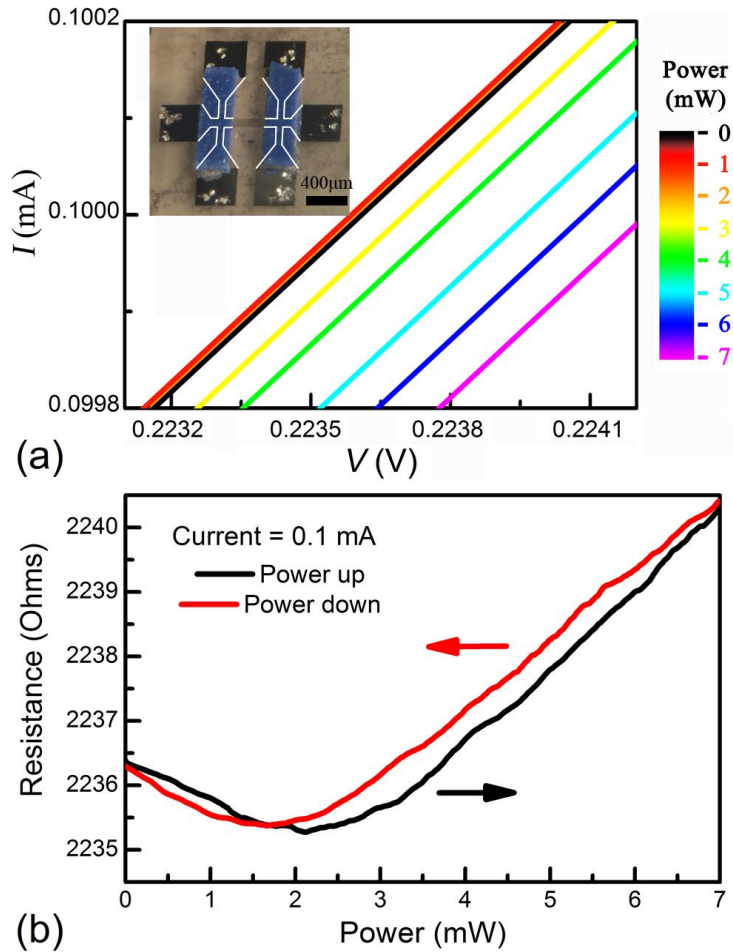

**Figure 2**. a) Power dependent optical gating effect on *I-V* properties under the illumination of blue light, with an optical image of the device shown in the inset.   b) Channel resistance as a function of light power with the current fixed at 0.1 mA and the power sweeping up and then down.

To verify this proposal, the expanded current-voltage (*I-V*) curves under optical gate operation at 10 K are displayed in Figure 2a. As expected, the *I-V* properties show a linear behavior based on the half-metallic property of LSMO under the blue light of different powers. With the light power rising from 0 to 7 mW, for the same voltage, the current increases slowly as the power increases to 2 mW and then decreases rapidly as the power increases further, implying the existence of a valley in the photon-induced change of the channel resistance. This valley is also observed in the channel resistance as a function of the light power in Figure 2b, which was measured with the current fixed at 0.1 mA and the power scanning up and down. The appearance of the valley can be explained by the competition between the thermal effect and the optical gating effect of the spin transistor which were both brought by the light illumination. Since the channel resistance decreases as the temperature increases due to the semiconducting behavior in this low temperature region, the increase of the light power would cause an increase of the sample temperature and thus a decrease of resistance. When the light power is low, compared to the optical gating effect which induces the resistance increase, the thermal effect plays a major role, leading to the reduction of the channel resistance. This tendency is confirmed by the up shift of the *I-V* curves for 1 mW and 2 mW. As the power increases further, the optical gating effect becomes dominant. Thus the resistance increase arising from the optical gating effect exceeds the resistance decrease induced by the thermal effect, giving rise to a fast increase of the resistance as the light power further increases. To confirm the spin origin of this optical gate operation, the contribution of the thermal effect is further excluded via the characterization of the power dependent *I-V* properties of a simple Hall device with the coverage in the spin transistor removed, where a monotonous decrease of the resistance was observed as the light power increases. Note that although a reduction of hole carriers would occur in LSMO under the light illumination,[26] an increase of the mobility due to the weakening of the ionized impurity scattering leads to the final decrease of the resistance. Therefore, an effective optical gate operation was realized for the control of the channel resistance via the modulation of the optical gate voltage. Note that, when the power sweeps up and down, these two curves in Figure 2b do not overlap with each other because of the temperature hysteresis to the power variation.

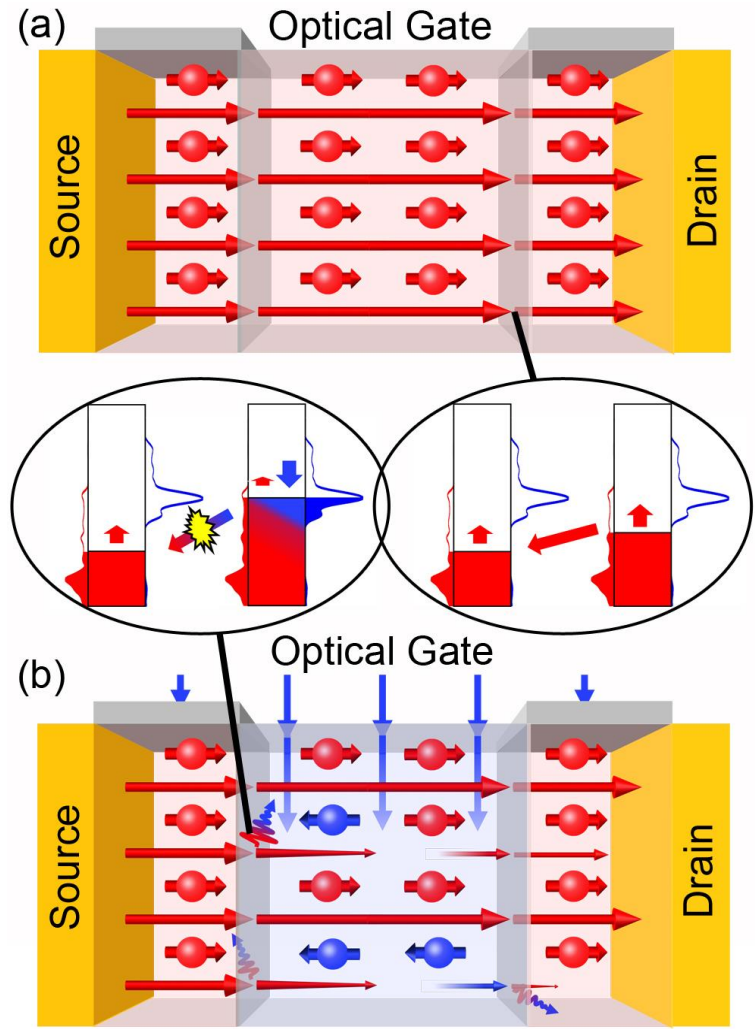

**Figure 3.** Schematics of photon-gated spin transistor under no light (a) or the blue light (b) with the DOS of the interface shown in the circle. Red rightward arrows with spheres stand for the major up spins in the whole channel under a rightward magnetic field in plane while the blue leftward ones denote the photon-induced down spins in the exposed area under illumination. Red and blue long arrows denote the spin-polarized current with different polarization.

In order to explain the optical gating effect in this device, a model of the photon-gated spin transistor was proposed as illustrated in **Figure 3**a, where the two terminals of the channel work as a source and a drain while the uncovered area works as an optical gate. When the optical gate is off, i.e. there is no light illumination, all the major up spins (as marked by red arrows with spheres) would remain along the channel in the same direction after an in-plane rightward magnetic field is applied and then withdrawn, in view of the half-metallic ground state of LSMO (the red filled area in the right panel of the DOS sketch). Under the voltage applied between source and drain, the Fermi level has a continuous change for adjacent areas. Thus a wholly spin polarized current forms and can easily pass the channel as shown by the red long arrows, meaning a low resistance gained under a zero optical gate "voltage". Nevertheless, the situation turns out to be dramatically different as the optical gate turns on. As a result of the photon-induced spin transition, many down spins (the blue arrows with spheres) appear in the area under the optical gate and then align leftwards under zero-field due to the tendency of the antiparallel alignment, which is opposite to the initial direction of the spin polarization in the current. In consequence, an obvious scattering occurs when the spin-polarized current meets opposite spins embedded in the photon-gated area, resulting in an increase of the channel resistance. This can be also explained by the mismatch of the DOS in the type of major spins between these two areas, as shown by the left panel of DOS sketch. In this scenario, the light illumination plays a critical role as an optical gate to modulate the channel conductance, based on which a photon-gated spin transistor works.

We now turn towards the cycling behavior of the photon-gated spin transistor. **Figure 4**a presents the variation of the channel resistance as the LED of blue light is switched off and on for five consecutive cycles with the corresponding LED states shown in the inset. The most eminent feature is the reproducible enhancement of resistance under light illumination, due to the photon-induced spin transition. Noteworthily, there is a sudden resistance peak when the light is on, which could be explained as follows: the photon-induced spin transition is an instantaneous behavior, hence the resistance of the spin transistor immediately increases to a larger value at the moment when the light turns on. After this transient variation, the channel resistance decreases somehow to an almost stable value because of the balance between the influence of the photon-induced spin transition and the light heating. When the light turns off, the resistance recovers slowly to its initial value after a sudden drop, which can be ascribed to the immediate disappearance of the photon-induced down spins but a slow cooling down process to the initial temperature.

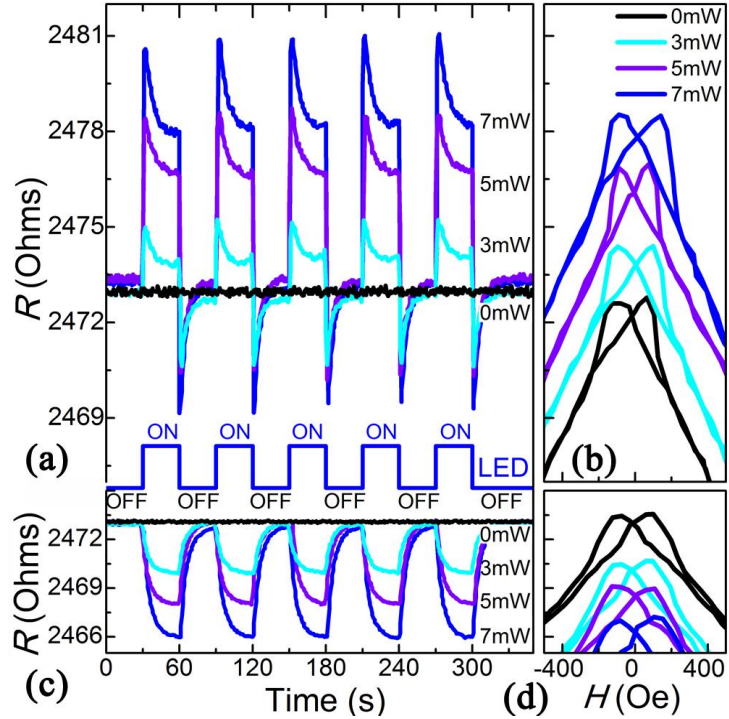

**Figure 4.** a) Cycling characteristic of the resistance ($R$) in the photon-gated spin transistor with the blue light of different powers switched off and on. b) $R$-$H$ curves of the photon-gated spin transistor for different light power. c) Cycling characteristic of the simple Hall bar recovering from the former spin transistor by removing the coverage, with the measurement performed under the same conditions. d) Corresponding $R$-$H$ loops of the simple Hall bar. The $y$-axis of resistance in (b) and (d) is in the same scale as that of (a) and (c) respectively. The inset between (a) and (c) depicts the LED states during the measurement of cycling characteristic.

Though the spin state returns immediately to the initial state (at 0 s) right after the light is off, the temperature is still high due to the light heating, so the resistance has a smaller value at the beginning and then recovers slowly to the initial value. Note that when compared with the time scale of the thermally induced resistance variation, i.e. about tens of second, the photon-induced variation of the resistance is in a scale of hundreds of millisecond, demonstrating a much faster speed. Whereas, this is not a limit in physics but a limit in the speed of the electrical triggering for LED light and the data recording time. As well known, for such a nonthermal photomagnetic effect involving the absorption of photons via electronic states, the variation is considered to be instantaneous (e.g. the rise time of the light pulse).[13] Furthermore, as expected from the power-dependent $I$-$V$ properties in Figure 2, the increase of the light power from 3 mW to 7 mW causes a larger variation of the resistance and the height of the stable value coincides well with the zero-point of the magnetoresistance measurement in Figure 4b, i.e., the resistance-magnetic field ($R$-$H$) curves, indicating that the stable resistance increases monotonously with increasing the light power from 3 mW to 7 mW.

In order to further confirm the key role of the optical gate in the photon-gated spin transistor, the insulating tape on the channel was removed and the spin transistor recovered to a simple Hall bar. Then the cycling properties were measured under identical conditions with corresponding data illustrated in Figure 4c. Much different from the variation trend of the spin transistor, the resistance of the simple Hall device does not show an instantaneous jump and drop at the moment of the light switching. Instead, there only exists a slow increase and decrease of the resistance which can be ascribed to the gentle temperature variation caused by the thermal effect. The reason is that the whole channel is exposed to light illumination inducing a simultaneous change of the DOS, so there is no discontinuity of the DOS along the channel and thus no intense scattering of the spin-polarized current, meaning no marked photon-induced transient increase of the resistance like the case of the spin transistor. Therefore, as the power increases, the only thermal effect becomes stronger and causes a larger decrease in resistance. Similarly, the stable value of the resistance in Figure 4c is consistent to the resistance at zero-field on the *R-H* curves in Figure 4d as the light power varies.

In this way, it is verified that the optical gate is indispensable in the structure of the spin transistor, while an effective repeatable gate operation of the channel conductance in the photon-gated spin transistor is demonstrated, which is tunable via changing the power of the light illumination. Note that this ON/OFF ratio is still small due to the limited variation in spin polarization, while the operation temperature is low because of the weak ferromagnetism and the low Curie temperature around 240 K, associated with a *M-T* behavior far away from the typical Curie-Weiss and Bloch's $T^{3/2}$ law. However, an improvement of the performance and the operation temperature can be expected in a magnet with a higher Curie temperature and a more prominent difference in spin DOS. By covering partial area of the whole channel, we can artificially create the source, drain and gate on the basis of the whole LSMO channel, circumventing the leakage problem of electric gate and the requirement for the efficiency of spin injection, which may promote the generation of more spin transistor devices. What's more, the instantaneous photon-induced change promises the possibility of the ultrafast operation in spin-FET for the application of data processing.

In conclusion, we proposed and prepared a new type of spin transistor with optical gating via partial illumination of the Hall bar channel under blue light of LED. Due to the photon-induced spin transition, the existence of light would cause a difference in major spins between the area under illumination and under protection and lead to an instantaneous increase of channel resistance, indicating an effective gate operation of the conductance via optical method. This optical gating effect has an obvious enhancement with the increase of the light power and shows repeatable cycling properties. Our findings would enrich the concept of spin-FET and promote the use of optical methods in spintronics for low power consumption and ultrafast data processing.

**Experimental Section**

*Sample preparation*: High-quality LSMO (20 u.c.) epitaxial films were grown on STO substrates by pulsed laser deposition from a target of stoichiometric $La_{1/2}Sr_{1/2}MnO_3$. The films was deposited at 650 °C with an oxygen pressure of 200 mTorr and then cooled down to room temperature under an oxygen pressure of 300 Torr to reduce the oxygen vacancy. A precise thickness of 20 unit cell at the atomic level was achieved according to the fine oscillating peaks of the in-situ RHEED (reflection high-energy electron diffraction).

*First-principles calculations*: First-principles calculations were performed via projector augmented wave (PAW) implementation of the Vienna *ab initio* simulation package (VASP). Taking the strong correlation in manganites into account, GGA+$U$ [27] was adopted as the energy function, where $U$ denotes the on-site Coulomb repulsion included for the d electrons of Mn. A setting of $U = 3$ eV and the exchange interaction $J = 0.98$ eV was used on the basis of previous works.[28,29] A $2 \times 2 \times 4$ supercell was used with atoms of La and Sr arranged alternatively, including 16 formula units with 80 atoms (Figure S3, Supporting Information). The value of the $SrTiO_3$ substrate was used for the in-plane lattice constant of a unit cell, while the experimental value gained from X-ray diffraction (Figure S1b, Supporting Information) was for the out-of-plane lattice constant. A $4 \times 4 \times 2$ $\Gamma$-centered $k$-point mesh was used for the structural optimization of atomic positions, while an $8 \times 8 \times 4$ $\Gamma$-centered $k$-point mesh was used for the calculation of DOS, with a cut-off energy of 500 eV for the plane-wave basis. Both the mesh and the cut-off energy were increased until convergence.

**Acknowledgements**

The authors thank Dr. J. B. Liu and National Supercomputing Center in Shenzhen for the help of first-principles calculations. This work has been supported by the National Natural Science Foundation of China (Grant Nos. 51322101, 51231004, 51571128 and 51671110) and Ministry of Science and Technology of the People's Republic of China (Grant Nos. 2014AA032901, 2014AA032904 and 2016YFA0203800)